# A probabilistic model to resolve diversity-accuracy challenge of recommendation systems

AMIN JAVARI

MAHDI JALILI[1]

___

Recommendation systems have wide-spread applications in both academia and industry. Traditionally, performance of recommendation systems has been measured by their precision. By introducing novelty and diversity as key qualities in recommender systems, recently increasing attention has been focused on this topic. Precision and novelty of recommendation are not in the same direction, and practical systems should make a trade-off between these two quantities. Thus, it is an important feature of a recommender system to make it possible to adjust diversity and accuracy of the recommendations by tuning the model.

In this paper, we introduce a probabilistic structure to resolve the diversity-accuracy dilemma in recommender systems. We propose a hybrid model with adjustable level of diversity and precision such that one can perform this by tuning a single parameter. The proposed recommendation model consists of two models: one for maximization of the accuracy and the other one for specification of the recommendation list to tastes of users. Our experiments on two real datasets show the functionality of the model in resolving accuracy-diversity dilemma and outperformance of the model over other classic models. The proposed method could be extensively applied to real commercial systems due to its low computational complexity and significant performance.

___

## 1. INTRODUCTION

In many e-commerce systems, users confront millions of options which they may purchase. Exploring such large item spaces is time consuming and is not manageable for most of users in many situations. Due to this reason, recommender systems have become a core component for many e-commerce applications and attracted many researches in recent years [1-3]. Although recommender systems work like an information retrieval system, there is no direct request for information in them. The major task of a recommender system is to use previous interactions recorded in the system in order to extract users' taste, and thus, provide a proper list of items to recommend. Recommender systems can be categorized into three classes based on the type of information used for recommendation: content-based, collaborative filtering and hybrid methods. Content-based recommenders work based on the content of the items and recommend items which have similar contents with those that a user has already purchased [4]. Algorithms based on collaborative filtering make recommendations for any user by the help of collective preferences of other users [5]. Hybrid methods use both the information available from

---

**Amin Javari**
Department of Computer Engineering, Sharif University of Technology, Tehran, Iran
e-mail: javari@ce.sharif.edu

**Mahdi Jalili**
Department of Computer Engineering, Sharif University of Technology, Tehran, Iran
e-mail: mjalili@sharif.edu

previous user-item interactions and contents of the purchased items [1, 6]. In many applications, content-based features are not easy to extract, and thus collaborative filtering approaches are the preferred ones.

Recommendation problem can be formulated as follows. There are a set of users *U*, a set of items *I* and a set of possible ratings *R* from *U* to *I*. Each user gives ratings on some items, which represents its utility value on the items. Since each user just experienced a small sector of the item space, the goal of a recommender system is to recommend a set of items for that user such that the user is likely to purchase them in near future and is satisfied by them. In some basic settings of the problem, the ratings have just values 0 or 1 that are extracted based on co-occurrence of users and items, i.e., if the user has purchased the item, the rating is 1, and 0 otherwise. For example, users' click on items in a website can be considered as co-occurrence. In some other cases, users represent their feedback on items through explicit ratings they give. Based on this framework, various recommendation algorithms have been introduced. One can mention correlation-based algorithms such as user-based and item-based collaborative filtering [7-9], spectral analysis techniques [10] and probabilistic models such as latent semantic model [11], Bayesian networks and Markov chain models [12, 13].

Collaborative filtering algorithm and its variants are the most frequently used recommendation algorithms in both academia and industry [14-16]. Many of collaborative filtering approaches do not take into account the sequence by which the items have been purchased by the users. The way users purchase items often has a sequential manner. For example, when a user watches a movie directed by a particular person and likes it, it is likely that the same user watches other movies from the same director. Markov chain (MC) based recommender is class of collaborative filtering based recommenders, which utilizes these sequential dependencies in their recommendations [16-19]. In MC models, the recommendation is modeled as a sequential prediction problem, and the goal is to predict items that a user will purchase in near future based on the user's last actions. To this end, first, a proper state space is defined and transition between states is estimated. Then, the recommendation for a target user is provided by applying the user's last actions on transition function. MC models consider the recommendation problem as a prediction problem and their task is to maximize accuracy of recommendation lists. However, systems which provide accurate recommendations do not necessarily satisfy users [20-22].

A practical recommender system should not only have a good accuracy but also provide proper novelty and diversity of items for the users [23, 24]. Novelty and diversity of recommendations describes the ability of the recommender models to suggest items which the users would not discover them by themselves [22]. In contrast to content-based algorithms which almost do not suffer from low diversity problem, most of the algorithms based on crowd preferences are biased towards popular items. This focus of recommendations on a small set of items happens since the classic recommendation models are designed to maximize precision of recommendations. Recommendation of popular items has lower risk in terms of precision [22]. On the other hand, regarding the diversity problem, we are interested to give same chance for all items to appear in the recommendation list. Thus, it could be argued that maximizing accuracy and diversity of the recommendation list are not in the same direction. Higher accuracy can be obtained by recommending some popular items; whereas personalized and diverse recommendations can be achieved by covering homogenously the whole items set without any focus on items with specific popularity interval. Thus, there should be a tradeoff between accuracy and diversity in recommendations [24]. Obviously, different frameworks may expect different levels of accuracy and diversity from their recommender systems. In other words, importance of accuracy and diversity may change in each framework. Therefore, capability of a recommender model to adjust diversity and accuracy level of recommendations is an important feature of a recommender system.

Given that MC-based recommender systems are model-based algorithms with proper functionality and low computational complexity [16, 19], in this paper, we introduce a novel probabilistic model that overcomes some drawbacks of classic MC-based methods. Furthermore, we propose a structure to approach the dilemma of accuracy-diversity by introducing a hybrid model, which can balance two recommendation algorithms: One specialized for high accuracy and the other one for high novelty. The individual algorithms are not optimal regarding to their diversity. By combination these two basic algorithms, we build a recommender system with not only good accuracy but also high diversity. The structure we introduce to resolve diversity-accuracy dilemma can be applied on all MC-based recommenders without increasing computational complexity of the models. Our results on two real datasets (Netflix and Movielens) show the effectiveness of the proposed algorithm in handling the accuracy-diversity challenge of the recommendation list as well as its superior performance over some classic algorithms.

## 2. Related works

Among different classes of collaborative filtering recommender systems, in this paper, we focus on MC-based models. Such models have many applications in analyzing users' behavior [25-28]. A number of works have applied Markov models in the context of recommender systems [17, 19, 29-31]. For example, Shani *et. al.* used Markov decision processes (MDP) for modeling the recommendation problem, by defining utility function for item set as a sequential optimization problem [13]. Garcin *et. al.* introduced a recommendation model based on evolving context trees, where context trees can be considered as variable order Markov models.

Although MC and MDP have been used in many real-world applications [17, 19, 29-31], there are some restrictions to employ them for recommendation. These models do not consider users' ratings on items. They do the recommendations based on the sequence of the items purchased by the user. The value of the ratings of the users is an important information resource which can be utilized to provide better recommendations. In this work, we utilize this information resource by defining a new state space. The diversity and accuracy dilemma has not been previously studied in MC-based models.

Novelty and diversity in recommender systems recently received much attention [21, 32-35]. It has been shown that performance of the recommender systems should be evaluated in terms of not only accuracy but also diversity and novelty [22, 36, 37]. Concerning this issue, different strategies have been introduced for diversification of recommendation lists [6, 38]. Typically, these models addressed the problem based on contextual or semantic information resources; for example, item-item similarities and user generated tags can be used to answer this problem [38, 39]. However, these models cannot be tuned to provide a desired level of diversity and accuracy and some of them also rely on more information resources. Recommendation algorithms may be used in different frameworks with different requirements. Flexible level of diversity and accuracy for a recommendation algorithm is an important capability for recommender systems. Zhou *et. al.* [24] proposed a model to address the challenge of diversity and accuracy dilemma which can be tuned to obtain arbitrary level of diversity and accuracy. Their recommendation process works based on the idea of heat diffusion across the bipartite users-items network. Although the model does not rely on any contextual or semantic information resources, their baseline model (HeatS) is not a widely used method. The model relies on a transition matrix between users and items. Real-world systems may contain millions of users which restricts scalability of the model for commercial applications. Apparently, the model proposed in [24] cannot be applied on other recommendation algorithms; it is indeed specific to the "HeatS" algorithm. However, our proposed models can provide a flexible level of diversity and precision for a widely recommendation algorithm. In this work, our baseline model is a probabilistic model based on Markov chains, and we introduce a structure that resolves the accuracy-diversity dilemma in this class of recommendations. Markov-based recommenders is a class of well-known and widely used techniques in real-world applications [16, 18, 19, 40]. The probabilistic structure introduced in this paper is flexible and can be applied on any Markov-based recommender.

## 3. Methods

In this section, we give the details of the proposed recommender system that is based on Markov model. The performance of this novel model is compared with three traditional recommender systems that have found stable application in commercial recommendation systems: memory-based collaborative filtering (CF), popularity-based recommendation algorithms and a classic Markov based recommender. Memory-based CF is one of the most successful techniques in recommender systems, which has been used in many commercial systems since introduction. This is mainly because of its simple implementation and stable inference.

### 3.1 Collaborative Filtering

In recommendation systems based on memory-based CF, in order to recommend a list of potential items to a target user, first, the model produces predictions for ratings of the target user on candidate items, and then, the items which have been predicted to receive high ratings from the user are recommender. Memory-based CF is often implemented in two forms: user based and item based. User-based CF works by, first, computing the similarities between the users based on pattern of their ratings to the items, and then, using these similarity values to predict the ratings on a target item. Item-based CF considers similarities between the items. In this manuscript, we take into account both these models. In order to compute similarities between the users, one can use metrics such as Pearson correlation coefficient or cosine similarity index. Using Pearson correlation coefficient, the similarity between users $i$ and $j$ can be obtained as:

$$S_{Pearson}(i,j) = \frac{\sum_{h=1}^{n}(R_{i,h}-\overline{R_i})(R_{j,h}-\overline{R_j})}{\sqrt{\sum_{h=1}^{n}(R_{i,h}-\overline{R_i})^2}\sqrt{\sum_{h=1}^{n}(R_{j,h}-\overline{R_j})^2}}, \quad (1)$$

where $R_{i,h}$ indicates the vote of user $i$ on item $h$ and $\overline{R_i}$ is average ratings made by user $i$. $n$ is the number of items rated by both users $i$ and $j$. In the same way, similarities between items can be extracted based on the ratings of the users on items [8].

By extracting similarities between users, predicted value for the rate of user $u$ on item $o$ can be obtained as:

$$P(u,o) = \overline{R_u} + \frac{\sum_{h=1}^{m}(R_{h,o} - \overline{R_h})S_{Pearson}(u,h)}{\sum_{h=1}^{m}|S_{Pearson}(u,h)|}, \quad (2)$$

where $S_{pearson}$ is the Pearson similarity of user $u$ with $h$, $R_{h,o}$ is the rating item $o$ received from user $h$ and $m$ indicates the number of ratings item $o$ received from different users. In item-based CF after extraction of similarities between the items, $P(u,o)$ can be obtained as:

$$P(u,o) = \frac{\sum_{i=1}^{n} Sim(o,i)R_{u,o}}{\sum_{i=1}^{n}|Sim(o,i)|}, \quad (3)$$

where $Sim(o,i)$ is the similarity of the item $o$ with item $i$ and $n$ is the number of ratings user $u$ gave to items.

## 3.2 Markov models

Unlike traditional CF algorithms, which only use recently purchased items for recommendation, methods based on Markov chain not only use that information but also take into account the information about the order in which users purchase items. Users often purchase items in a sequential manner such that a particular item is purchased first, and then, another item is purchased and so on. Markov models use this information and give different weights for the items based on the order in which they have been purchased by users.

In order to employ Markov chain for recommendation, appropriate state space and transition function should be first defined. State of each user can be represented by the set of the items which have already been purchased (or rated) by the user. It is well-known that in order to have an acceptable performance for a recommender system, the constructed connectivity matrices should be sparse [5]. Different users may purchase different number of items and due to the sparsity problem, only a sequence of at most $m$ items is considered as a particular user's state [13]. Let us consider the vector $S_u = <I_m, I_{m-1},...,I_1>$ as state of user $u$, which denotes the user's last $m$ purchased items in a sequential manner. As the state space is defined, the transition function should be estimated. The transition function describes the probability that a target user that has purchased items $I_m$, $I_{m-1},...,I_1$ will select item $I_{m+1}$ in the next step. According to the train dataset which consists of different users' ratings on the item and based on maximum likelihood estimation, the transition function between two states can be estimated as:

$$TF(<I_m,I_{m-1},...,I_1>,<I_{m+1},I_m,...,I_2>) = \frac{N(<I_{m+1},I_m,...,I_1>)}{N(<I_m,I_{m-1},...,I_1>)}, \quad (4)$$

where $N(<I_m,I_{m-1},...,I_1>)$ indicates the number of users visiting the state $<I_m,I_{m-1},...,I_1>$ in their rating sequences in training dataset.

## 3.3 State automata-based recommender system

The recommendation method proposed in this paper, similar to the one based on Markov chain, uses the sequential manner of users' ratings. However, in our proposed method, we define the states and the prediction model in a way different from previous Markov chain models. Furthermore, we introduce two prediction models to solve accuracy and discovery dilemma in recommender systems. It is well know that accuracy and diversity of the prediction list are not in the same direction; often, when accuracy improves, the diversity of the recommendation list worsens and vice versa [22]. To the best of our knowledge, there is no Markov-based recommendation algorithm that gives the full power to control the accuracy and diversity of the recommendation list. Here, we introduce a hybrid model that has a single parameter controlling desired levels of accuracy and diversity of the system. We separate the proposed method into two parts. In the first part of the algorithm, two primary graphs are extracted from training dataset. In the second part, the recommendation is performed based on two separate prediction models. In the next sections, we articulate each part in details.

### 3.3.1 Extraction of aggregated transition graph and aggregated co-occurrence graph

By considering each item as a state, users' sequential ratings can be modeled as transition between different states. However, in order to include the value of ratings in the transition probabilities, we model each item by two states: positive (*like*) state and negative (*dislike*) state which we indicate in the rest of the paper by L-state and D-state, respectively. L-state indicates that the target user rates the item positively, whereas negative ratings are indicated by D-state. Let us denote L-states and D-states of item $i$ by $s_{i,L}$ and $s_{i,D}$, respectively. In other words, we model the item space in $2I$ states and users' ratings as transitions between these states where $I$ is number of items. Each edge $(s_{i,L}, s_{j,D})$ in this transition graph represents consecutive ratings. For example, ratings of a specific user in table 1 can be transformed to state transition model shown in figure 1.

| Item ID | Rate | Time Step |
|---|---|---|
| 23 | Dislike | 1 |
| 532 | Like | 10 |
| 43 | Like | 21 |
| 389 | Dislike | 23 |

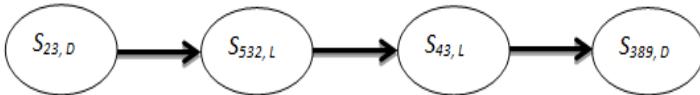

Figure 1: Rating history of a specific user (Table) and the state transition model of the same user's ratings (Figure).

In order to aggregate the information about users' behavior and their sequential ratings, we use statistics extracted from the transitions in the state space. In this way, we introduce an aggregated transition (AT) graph, which is a state space with $2I$ states. In this graph, the weight of connection between two states $s_{i,L}$ and $s_{j,D}$ is defined as the number of users who have the edge $(s_{i,L}, s_{j,D})$ in their transitions. Indeed, we count the number of users who visit item $j$ after item $i$ and have similar tastes on these two items (i.e., all liked item $i$ and dislikes item $j$).

Essentially, in order to implement our proposed model, the datasets should have a time tag for all ratings. For datasets without such labels, we define aggregated co-occurrence (AC) graph. In this work, we define weight of the connection from state $s_{i,L}$ to state $s_{j,D}$ as the number of users who have rated both items $i$ and $j$ and given similar ratings to them regardless of rating sequence. Although in this way we lose some information about the rating sequence, it makes the algorithm a general one and makes it possible to compare with other algorithms that do not consider the rating sequences (e.g., user-based CF). AC graph suffers less than AT graph in terms of sparsity problems. The sum of the edge weights in AT graph is $N$, where $N$ is number of ratings in training dataset, whereas AC graph, in average, has the sum of edge weight as $U\bar{N}^2$, where $U$ is number of users in training dataset and $\bar{N}$ is the average number of ratings for each user. By considering $\bar{N}$ as $N/U$, the sum of edge weight becomes $N^2/U$, and thus, the sparsity is much less in AC graph as compared to AT graph.

### 3.3.2 Recommendation

Based on the aggregated transition graph introduced above, in this section, we introduce two probabilistic recommendation models. One of the models aims at providing a recommendation list with high levels of precisions, while the other model tries to recommend items that have the best fit to the target user's taste. In order to perform the recommendation based on these probabilistic models, we first extract the state of the target user. We define state of each user as its previous ratings on the items. As mentioned above, users' ratings can be considered as nodes in AT or AC graphs. Thus, the state of the users is a vector of nodes' identification (ID) numbers. In other words, the state of users consists of some sub-sates where each node in AT or AC graphs indicates a sub-state. For example, state of user with ratings as indicated in Figure 1 can be defined as <23D, 532L, 43L, 389D>. Unlike traditional Markov models, we do not consider sequence of ratings in the vector of users' state. The definition we have considered for the state of users has high flexibility. Using this method, it is possible to filter out some certain nodes with special characteristics in defining the state for a specific user. It is also possible to set the state of a user as weighted vectors of sub-states. For example, the items purchased recently might be more informative about taste of the users than those purchased previously; our definition allows intensifying the weights of nodes that belong to recently purchased items.

*Recommendation based on precision maximization*

In traditional prediction models based on Markov chain, the purpose is to recommend items that are likely to be purchased in near future by the target user. Here, we introduce a similar method that not only predicts the items which the user is likely to purchase them in future but also predicts the items that the user will like them. The classic Markov chain models take into account only information about users' purchased items, but our model also considers the value of ratings. This is performed with the help of separating each item into two distinct states. In order to provide the recommendation list, we first produce probabilities about users' interest on each item, and then, we recommend items with the highest probability values. The probability of the interest of user $u$ on item $i$, $P(I_i/S_u)$, is obtained as:

$$P(I_i \mid S_u) = P(s_{i,L} \mid S_u) - P(s_{i,D} \mid S_u), \qquad (5)$$

where $S_u$ is the state of the user $u$, $P(s_{i,L}/S_u)$ is the probability of transition from the state of user $u$ to L-state of item $i$ and $P(s_{i,D}/S_u)$ is the probability of transition to D-state of the item $i$. According to this equation, recommendation score for item $i$ is the probability that user *u* likes this item minus the probability that it dislikes the item. Indeed, the purpose of this model is to recommend items with the highest predicted value of popularity and the lowest predicted value of being uninterested. The probability of the interest of user *u* in item *i* can be obtained as:

$$P(s_{i,L} | S_u) = \sum_{k=0}^{m} P(s_{i,L} | S_u(k)) \times P(S_u(k)), \qquad (6)$$

where *m* denotes the size of the state vector for user *u* and $P(S_u(k))$ indicates the weight of sub-state *k*. $P(S_u(k))$ is a tunable parameter that can be optimized to have good performance. It this work, we did not aim at optimizing this parameter and set the value of $P(S_u(k))$ for all sub-states equal to $1/m$. $P(s_{i,L}/S_u(k))$ is the probability of transition from sub-state *k* of user *u* to $s_{i,L}$. This transition probability can be extracted using maximum likelihood estimation method. Based on the maximum likelihood estimation, the transition probability between node *i* and *j* is the weight of the edge (*i*,*j*) in AT or AC graphs divided by out-degree of node *i*. $P(s_{i,D}/S_u(k))$ can also be obtained in the same way.

Indeed, equation (6) indicates the probability that user *u* will like item *i* (i.e., will transit to the sub-state $s_{i,L}$ in the next step), if user *u* has visited a vector of sub-states, $S_u$. To find the transition probability of user *u* to the sub-state $s_{i,L}$, we first obtain the probability of transition (dependency) from sub-states visited by user *u* to $s_{i,L}$, and then aggregate the obtained values in a weighted manner. We obtain the probability of transition (or dependency) between the sub-states based on the users' behavior in the training dataset. The weight of each sub-state indicates the importance of the sub-state in describing the behavior of the target user. For example, in a new recommendation framework, we may define the weights of sub-states (i.e., news items visited by the target user) based on the time the user spent on the news. As the user spent more time on a news item, it means that the item better describes the user's interests. However, in this paper, we do not consider this issue and set the weights of sub-state equal to $1/m$.

In order to provide the recommendation list, we first, compute the above probabilities for all candidate items and then recommend top-*L* items with the highest predicted values, where *L* is the basket size of the recommendation list. Although this prediction model provides a list with good precision, it often has a tendency to recommend high-degree and popular items. This makes the model suffer from novelty and diversity problems, i.e., the recommendation list does not often provide diverse and novel items to the users. Let us denote this model as PM model.

In order to show that PM model is biased toward recommending high-degree items, we model each item with one state and extract the item's transition function regardless of the rating values. Like most of real datasets, let us consider the item space consists of items with different popularities (e.g., in-degrees). If we obtain the expected value of the target user's transition to different items, it is easy to show that the items with higher popularity would gain higher expected value and the items' probability to be recommended would be related to their popularity. Thus, in general, PM model is often biased to select items with low novelty (high degree). Next, we introduce another model, which takes into account the novelty and diversity of the recommendation list.

*Recommendation based on specification maximization*

The model we introduce in this section tries to recommend items that are specific to the target user. This model extracts the recommendation list in a different way than PM model and tries to complement it. Let us denote this model as specification maximizer (SM) model. The goal of SM model is to find a list of items that is specific to the target user with the hope that maximum user satisfaction is obtained. SM model, first, extracts specification probability for all candidate items with respect to a target user and then recommends items with the highest probability values. The probability of specification can be obtained as:

$$P(S_u | I_i) = P(S_u | s_{i,L}) - P(S_u | s_{i,D}), \qquad (7)$$

where $P(S_u/I_i)$ indicates that if we recommend item $i$ to user $u$, to what extent the interest in item $i$ is specific to user $u$. This probability can be obtained based on two other probabilities: $P(S_u/s_{i,L})$ and $P(S_u/s_{i,D})$. $P(S_u/s_{i,L})$ can be computed as:

$$P(S_u | s_{i,L}) = \sum_{k=0}^{m} P(S_u(k) | s_{i,L}) \times P(S_u(k)), \qquad (8)$$

where $P(S_u(k)/s_{i,L})$ is the probability of conditional transition from $S_u(k)$ to $s_{i,L}$. Supposing that user $u$ has been visited $s_{i,L}$, $P(S_u(k)/s_{i,L})$ indicates that in which probability the previous sub-state visited by $u$ is $S_u(k)$. According to this equation, when an item is relevant to the state of a target user and unpopular for other users, it would have higher chance to be recommended to this user. It could be discussed that equation (8) indicates that to how much extent item $i$ is specific to user $u$. Personalized recommendations for each user contains items which are relevant to the target user and are less relevant for the users with different profiles that user $u$. In other words, items that are common interests for users with different profiles are not good candidates for personalized recommendations. In general, low-degree items provide higher levels of specification. Therefore, we expect SM model to give higher scores to items with lower degrees.

Transition function for SM model can be extracted from AT or AC graphs based on maximum likelihood estimation method. Thus, in SM model, the transition function from node $i$ to $j$ is defined as the weight of the edge $(i,j)$ in AT or AC graphs divided by the in-degree of node $j$. Compared with PM model, SM model provides a recommendation list that is biased toward low-degree items. It is intuitive that, often, items with high degrees provide less specification than low-degree items. Items with high popularity (e.g., degree) have better chance than those with low popularity to be attended, and thus, less popular items are likely to be specific to users.

Let us further explain PM and SM models by an example. Suppose that item $i$ has high dependency on the profile of user $u$ and the most of the users with similar tastes as user $u$ have purchased item $i$. Therefore, based on equation (6), the item would get a high score of recommendation for user $u$. On the other hand, suppose that users with different interests as compared to user $u$ also have high probability of transition to this item, which means that the interest in item $i$ is not specific to the profile of user $u$. Therefore, based on equation (8), this item would get a low score of recommendation. In an example from movie recommendation framework, consider a movie which wins an Academy Award. In general, a large group of users are probable to visit this item in the future. Therefore, recommending this item is favorable regarding the precision criterion. However, since this item is a common interest of users with different profiles, recommending the item would not satisfy the personalization.

We investigate structure of SM and PM models in an example shown in Fig. 2. Let us consider two sets of states A and B which form a small state space. Weight of the edges between any $A_i$ and $B_j$ is the number of users that are transmitted from $A_i$ to $B_j$. Based on these weights and maximum likelihood estimation, probability of transition for a user from $A_1$ to $B_1$, $B_2$ and $B_3$ in the next step is 0.80, 0.20 and 0, respectively. Thus, the highest transition probability is from $A_1$ to $B_1$. It is also possible to determine the most probable origin. Let us suppose we are at state $B_1$, B2 or $B_3$. Based on maximum likelihood estimation strategy, the probability that the user in the previous step was started at $A_1$ is 0.40, 0.5 and 0 for $B_1$, B2 and $B_3$, respectively. Thus, $A_1$ is the most probable start point if the state is $B_2$. SM and PM models perform in the same manner as this example. PM model recommends items, which a user $u$ has the highest probability to transmit their L-state from $S_u$. Whereas, SM model recommends items, which if the user stated at its L-state (i.e., if we recommend the item to the user and he/she becomes satisfied with the item), it has the highest probability that the user was transmitted from $S_u$.

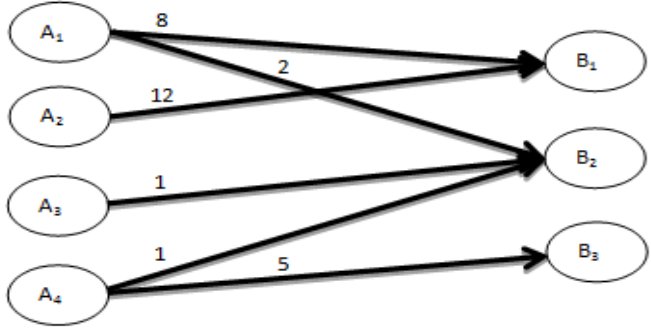

Figure 2: A Sample bipartite state space (weights of the edges indicate the number of users that are transmitted from origin state to destination state).

If we investigate PM and SM models based on the popularity of items in their recommended lists, in average, both of them recommend items which are relevant to the target user; however, PM model recommends items with almost high degree (popularity), while SM model introduces items with almost low degree. It is clear that items with high degree are likely to attract users' attention in near future, which would result in high precision for the model. On the other hand, this type of recommendation often results in low novelty and can be considered as obvious recommendations. Unlike to PM, SM recommends the items with low degree which increases the risk of recommendations and makes the lists more informative. We introduce a hybrid model in the next section in order to integrate these two models to construct a recommender with controllable levels of precision and novelty.

*Hybrid recommender*

We articulated that SM and PM models provide the recommendation lists by producing a vector of probabilities on items set. Here, we introduce a hybrid model, which uses both SM and PM model aiming at providing a recommendation list that is user-specific with good precision as much as possible. The hybrid model is simple model that linearly combines the outputs of SM and PM models. The probability vector of our proposed hybrid model, $P_{CP}$, is obtained as:

$$P_{CP} = \alpha\, P_{SM} + (1-\alpha)\, P_{PM}, \qquad (9)$$

where $P_{SM}$ and $P_{PM}$ are the probability vectors extracted by PM and SM models, respectively. $\alpha$ is a parameter in the range [0, 1] that is tuned by the system owner depending on the performance he/she is looking for. As *α* increases, the final recommendation list has a tendency to SM model, whereas small value of *α* dominates the PM model and gives more weight to the precision of the recommendation list.

SM and PM models could be combined in other ways. For example, items can be ranked based on $P_{SM}$ and $P_{PM}$, leading to a ranking vector obtained from each of the models which we call them $R_{PM}$ and $R_{SM}$. The vector of the rankings can be linearly combined as

$$R_{CP} = \alpha \, R_{SM} + (1-\alpha) \, R_{PM}. \tag{10}$$

The obtained vector, $R_{CP}$, could be used to do the final recommendations.

### 3.3.2 Structural analysis of the proposed model

One of the main problems of Markov chain model is how to determine the size of the state. By increasing the size, and due of the sparsity problem, the extracted transition function would become unreliable. On the other hand, small number of chains is not able to appropriately represent users' state. Since a user's states in such models consider only the previously purchased items by the user, it does not take into account user's ratings value. Also as mentioned, models based on Markov chain often result in good precision by recommending popular items, while its novelty is often low. Our proposed model tries to tackle these challenges by new definition for the state of users and modifying the prediction model. Since state vector of each user consists of all items purchased by the user, we do not ignore any data in the state definition and the states – as weighted vectors – provide flexibility for the model to add more information about each item. For example, it is possible to give higher weights for items recently purchased by the user. Also, separating each item into two states makes our model be able to take into account rating values.

The most significant contribution we have in our proposed model is to solve the novelty problem of traditional Markov models by introducing a hybrid model which is able to address the novelty and precision challenge by tuning a hybridization parameter. Markov-based methods are model-based algorithms for recommendation systems which have computational complexity of $O(n)$ where *n* is the number of ratings in the training dataset. Compared with other recommendation algorithms, they have considerably better computational complexity. Also, these models outperform other traditional models in terms of precision. However, the main problem that restricts the usage of Markov models in commercial systems is low novelty of their recommended lists. Our proposed model has the same computational complexity with the original Markov model, while the novelty of its recommendation list is much higher.

## 4. RESULTS

In order to compare the performance of the proposed recommendation algorithm with that of some classic algorithms, we applied them on Netflix and Movielens datasets, which are frequently considered as benchmarks in recommendation problems.

## 4.1 Datasets

Two datasets (Netflix and Movielens) were employed to evaluate performance of the proposed method. Both datasets are ratings of users to set of movies on a scale of 1-5. The Netflix dataset used in the experiments is a random subset of the original Netflix dataset and is with 9,983 users, 6,533 items and 2,041,247 ratings. Movielens dataset is with 6,040 users, 3,952 items and 1,000,209 ratings. Table 1 summarizes the statistics of the datasets. For Netflix dataset, only the date of the ratings is available, while for Movielens dataset, the time label of ratings is available in seconds. According to the structure of our model and since in many applications explicit ratings are not available, we transformed rating values from scale of 1-5 to ratings with *Like* or *Dislike* form. In order to do this, we defined value of 2.5 as a threshold and considered ratings above the threshold as *Like* and ratings below the threshold as *Dislike*.

Table 1: Statistics of Netflix and Movielens datasets used in this work.

|  | Number of users | Number of items | Number of ratings |
|---|---|---|---|
| Netflix | 9983 | 6533 | 2041247 |
| Movielens | 6040 | 3952 | 1000209 |

## 4.2 Performance Metrics

To test recommendation models, we divided the datasets into two parts: training and test datasets. Datasets were sorted based on time label of ratings, and the first 90% of the ratings were considered as training dataset and the rest as test dataset. In order to test the recommendation algorithms, we used users who had at least one rating in the training dataset and their number of ratings in test dataset was above the size of the recommendation list. Among five measures we employed to test the performance of the algorithms, four of them concern the recommended lists with size $N$ and one concerns the whole items ranked by the recommendation algorithm. In our experiments, we used an off-line evaluation. For each test user, we hided test dataset and asked the recommender algorithms to perform the recommendations based on the user's ratings in training dataset. The quality of the recommendations was evaluated based on the items the test user has already purchased in the test dataset.

### 4.2.1 Recovery

In order to evaluate performance of recommender algorithms in giving a proper ranking to whole item set, we employed the recovery metric. We prefer systems that give higher rank for items that are relevant to the target user. Relevant items to each user can be extracted based on her/his ratings in test dataset. We considered items purchased by a test user and get *Like* rating in the test dataset as relevant items to the target user. Hence, recovery $R$ can be obtained as:

$$R = \frac{\sum_{u \in u_{TestSet}} \frac{1}{L_u} \sum_{i=1}^{L_u} \frac{r_i}{C_u}}{|u_{TestSet}|}, \qquad (11)$$

where $C_u$ is the number of candidate items for recommendation in item set, $L_u$ is the number of relevant items to user $u$, $r_i$ is the place for item $i$ in the ranked list for user $u$ and $|u_{TestSet}|$ is the number of users in the test dataset. According to this definition of recovery, the lower $R_u$ is, the more accurate the system.

### 4.2.3 Precision

Many applications are designed so that they recommend $N$ items to users. Precision for the list recommended to user $u$, $P_u(N)$ is defined as the percentage of the relevant items to user *u* in the list recommended to the user. We considered items purchased by the target user in test dataset and received *Like* rating as relevant items to the target user. Precision of the systems on a recommendation list with *N* items can be defined as:

$$P(N) = \frac{\sum_{u \in u_{TestSet}} P_u(N)}{|u_{TestSet}|}. \qquad (12)$$

### 4.2.3 Item space coverage

A good recommendation method supports large part of item space in its recommendation list. In other words, we are interested in methods which can recommend large proportion of items in their recommendations. To measure this property, the percentage of items that are recommended to test users can be extracted. However, this measure does not take into account the number of times each item appears in recommendation lists. Some items may appear in the lists frequently, whereas some others might be recommended only once. In order to have a metric without such a problem, item space coverage is given by Shannon entropy as:

$$C(N) = -\sum_{i=1}^{L} p(i) \log p(i), \qquad (13)$$

where $p_i$ is the percentage of recommendation lists that contains item *I* and *L* is the number of candidate items.

### 4.2.4 Diversity

One of the measures which concerns about personalization in recommender systems is inter list diversity measure. It is desirable in recommender systems to recommend an item set which is unique for the target user and fits her/his interests. For two users *i* and *j*, the distance between their lists ($d_{i,j}$) can be obtained as:

$$d_{i,j} = 1 - \frac{c_{i,j}}{N}, \qquad (14)$$

where $c_{i,j}$ is the number of common items in the lists recommended to these users, and *N* is size of the recommended lists. Inter list diversity *D(N)* is the average distance (as defined above) between all test users. As *D* is higher, the method recommends more personalized recommendation lists.

### 4.2.5 Novelty

A recommendation list becomes more informative and novel as the target user is less likely to know the existence of the items form the recommended list. Various methods have been introduced to evaluate novel recommendations [37]. Self-information based novelty is a measure for novelty relative to popularity of items [33]. According to this measure, popular items provide less novelty. Self-information based novelty can be obtained as:

$$S(N) = \frac{\sum_{u \in u_{TestSet}} \left( \sum_{i=1}^{N} \log_2(\frac{|u|}{rels_i}) \right)}{|u_{TestSet}|}, \qquad (15)$$

where $|u|$ is the number of test users and $rels_i$ indicates the number of users that have purchased item $i$.

## 4.3 Experimental results and discussion

The proposed hybrid model has a parameter $\alpha$, which largely influences its performance. This parameter can be customized depending on the desired performance. In our experiments, we evaluated performance of the hybrid model as a function of hybridization parameter $\alpha$. We compared the performance of the proposed model with that of three well-known recommendation algorithms: memory-based CF (item based CF and user based CF), classic Markov model and popularity-based recommender. We also compared two forms of the proposed algorithm: time aware (known time tags of the ratings) and time unaware (unknown time tags).

**4.3.1 Dependence of the performance measures on α**

Figure 3 shows the performance of the proposed method (as compared to user-based CF and popularity-based method) in terms of different metrics in Netflix and Movielens datasets. In both datasets in sequence aware (SA) version of the model, the value of $\alpha = 0$ (PM model) results in the highest precision and the lowest novelty, whereas the value of $\alpha = 1$ (SM model) results in the highest novelty and the lowest precision. These two metrics have a monotonic performance (decreasing for precision and increasing for novelty) as $\alpha$ increases from 0 to 1. However, in Sequence Unaware (SU) form of the model in both datasets $\alpha = 0.1$ results in highest precision. Depending on the performance one would like to have (precision might be more important in some application and less in some others), $\alpha$ can be tuned so that the desired performance is obtained. In our experiments, we could always find a value of $\alpha$ such that both precision and novelty of the proposed recommendation algorithm is higher than user-based CF and popularity-based methods. In terms of recovery metric, both SM and PM algorithms give higher ranks to items that are relevant to taste of the target user, with recovery rates always below 0.5. Note that the value of the recovery rate as 0.5 for a recommendation algorithm indicates that the algorithm works like a random machine in producing ranked list of items for a target user.

Given the time series of items' popularity, their popularity in near future is predictable [41]. It can easily be shown that items with low popularity are unlikely to become popular in a short time horizon such as 1-2 weeks. Hence, recommendation of popular items has a low risk in terms of precision measure, while recommending unpopular items increases the risk of hit ratio and often results in low precision. In order to have good precision of recommendation, users should purchase items with the highest popularity more than those with lower popularity. As mentioned, SM favors items with low degree, while PM is biased toward items with high degree. These specifications of SM and PM models justify the higher precision and also better recovery rate of PM model as compared to SM model.

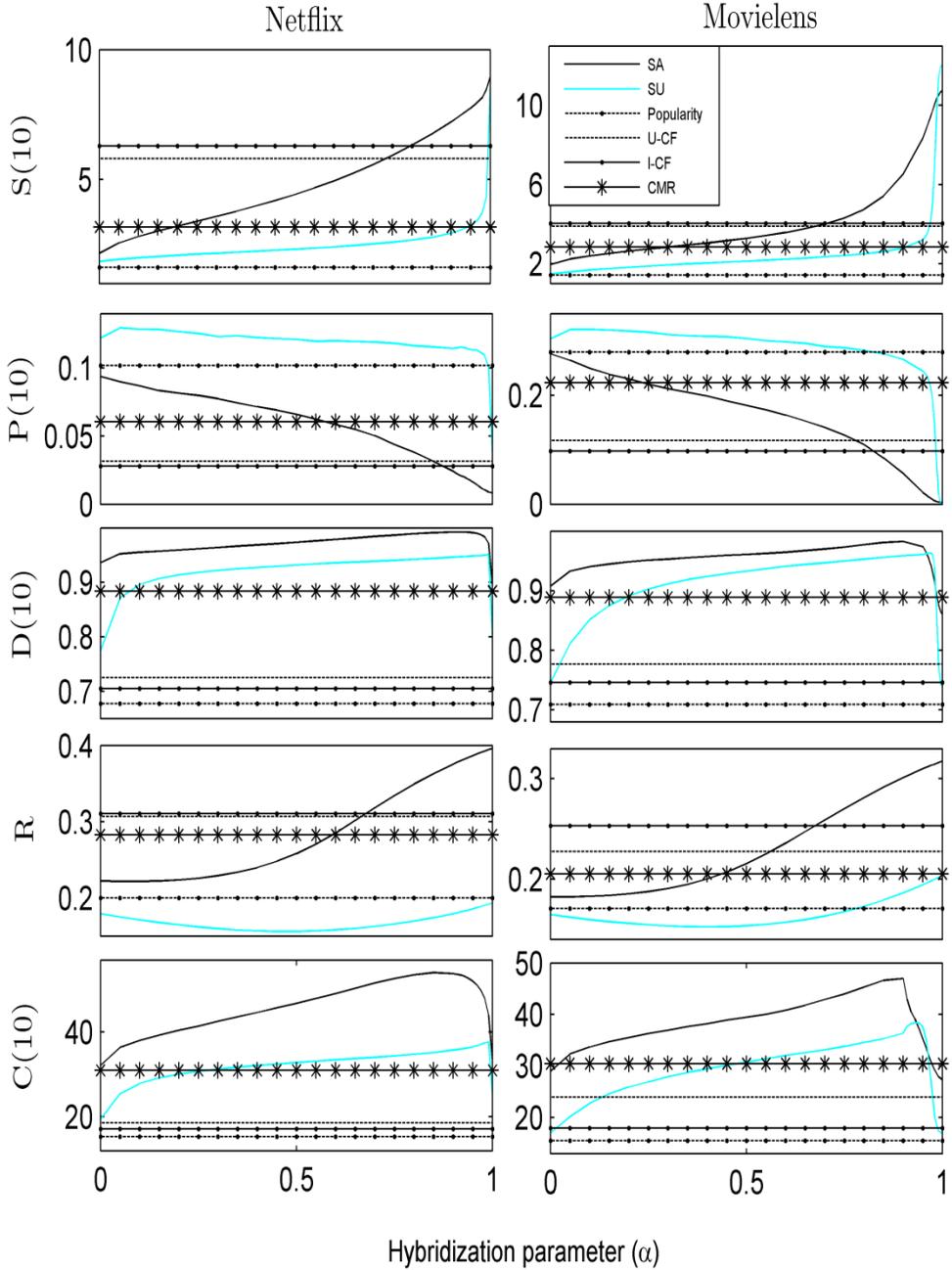

Figure 3: Performance of the proposed recommendation method (sequence aware hybrid (SA hybrid) and sequence unaware hybrid (SU hybrid)), user-based CF (U-CF), item-based CF (I-CF), popularity-based recommender (Popularity) and classic Markov recommender (CMR) as a function of hybridization parameter $\alpha$ on Netflix and Movielens datasets. The size of recommendation list is set as 10 and the performance of the methods is evaluated based on novelty S(10), precision P(10), diversity D(10), recovery R and coverage C(10).

The proposed hybrid recommendation method has by far better performance than memory-based CF models (both item-based and user-based forms), and popularity-based methods in terms of diversity and coverage measures. Also, there is optimal value of $\alpha$ for these metrics. In other words, for the optimal value of $\alpha$, the proposed hybrid model has better coverage and diversity than individual PM and SM models. Each of PM and SM models has focus on only particular part of item space and their recommendations partially cover the item space. This restriction causes these models to have low coverage. The hybrid model overcomes this restriction and makes it possible to cover whole item space in recommendations. Indeed, the proposed hybrid model considers items with different degrees in recommendation while SM supports items with low degree and PM supports high-degree items. Figure 4 shows the degree distribution of recommendations of hybrid model for different value of $\alpha$ on Netflix and Movielens datasets. The proposed model with $\alpha = 0.7$ and $\alpha = 0.4$ gives chance to items with different degrees to appear in recommendation lists, where $\alpha = 0$ or $\alpha = 0.9$ only consider items with special degree characteristics in their recommendations. Our results show that for both datasets, diversity and coverage measures reached their maximum for $\alpha \sim 0.9$.

Figure 5 demonstrates the degree distribution of Movielens and Netflix dataset. As the distributions show, about 20 percent of the items attract 80 percent of users' ratings in both datasets. Thus, a large number of items in the item space have low degrees and, thus, low popularity values. When we focus on items with low degree ($\alpha \sim 0.9$), the model covers large number of items in its recommendations and the coverage becomes maximum. However, the bias toward very low-degree items for the values $\alpha > 0.95$ causes the model to lose coverage and diversity. Overall, hybridizing SM and PM models not only is a mean to resolve accuracy and novelty dilemma, but also significantly improves diversity and coverage as compared to individual SM and PM models as well as user-based CF and popularity-based recommendation methods.

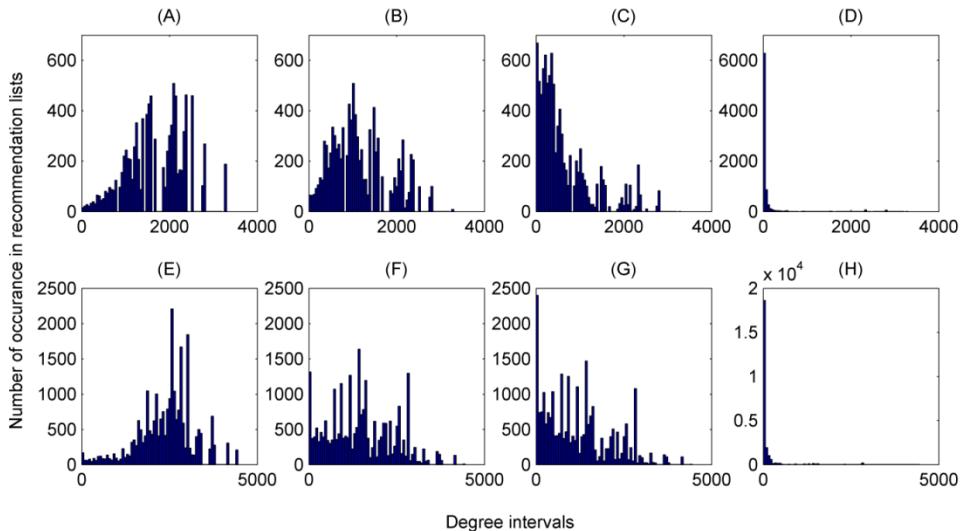

Figure 4: Number of occurrence for items with different degree (popularity) values in the proposed hybrid model. A, B, C and D represent the histograms of degree of recommended items for values

of $\alpha$ as 0, 0.2, 0.7 and 0.95, respectively, on Movielens dataset. E, F, G and H represent the histograms for values of $\alpha$ as 0, 0.2, 0.4 and 0.9, respectively, on Netflix dataset.

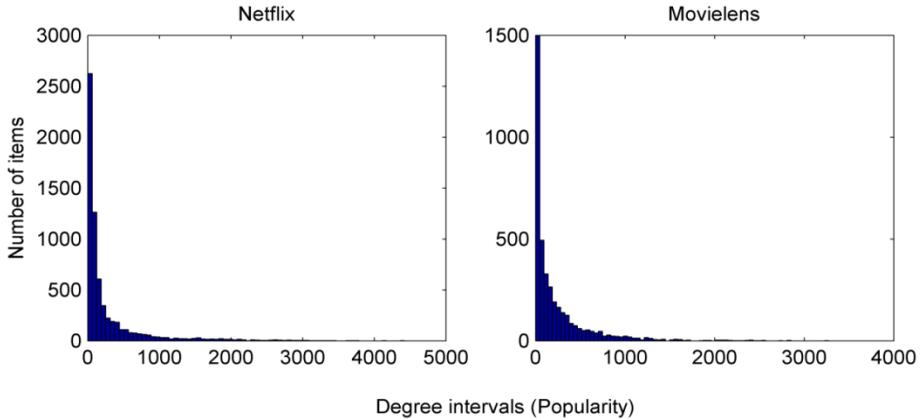

Figure 5: Degree distribution of items in Netflix and Movielens datasets.

To better analyze the behavior of the model as a function of the hybridization parameter, let us give a toy example. Suppose that we have three classes of items with different popularity intervals: very unpopular (A), unpopular (B) and popular (C). These three classes contain 20%, 60% and 20% of items, respectively. Intuitively, we may say that SM model gives higher score for items that are relevant to the target user and also belong to class A. PM model has a bias toward items that are relevant to the target user's taste and belong to C. Thus, in general, PM model recommends items in class C and SM covers items in class A; each of these models eliminate about 80% of the item space in their recommendations. As we mentioned, hybridization of PM and SM models overcomes this restriction and the hybrid model considers the items from the all three classes in its recommendations.

#### 4.3.2 Sequence aware (SA) and sequence unaware (SU) forms of the model

In general, SA and SU forms of the proposed recommendation algorithm have almost similar behavior as a function of $\alpha$. SU form of PM model as compared to SA form of the same model is more biased toward high-degree items, whereas SU form of SM model is more biased toward low-degree items than SA form of the model. Due to this property of SA models, SU form of PM model provides higher precision compared to SA form of PM model. Similarly, higher novelty can be achieved by SU form of SM model than SA form of the same model. Related to this characteristic, SU form of our proposed hybrid model has lower diversity and coverage. This is due to the fact that the sector of item space that is covered by SU model is smaller than the one covered by SA models.

#### 4.3.3 The proposed model versus classic baseline models

In the implementation of user-based CF, we used weighted average of 50 users that have rated the target item and have had the highest similarity with the target user. Also, in the classic Markov-based model we used Markov chain of order 2. Popularity-based model recommends items to the target user which has the highest popularity among those items that have not yet purchased by the user. Although popularity-based recommender has proper performance on precision and recovery parameters and outperforms the proposed hybrid models for some value of α, it has low novelty. Moreover, since the popularity-based model only covers items with high popularity, its performance in terms of coverage and diversity measures is low as compared to other models. This makes this method have little applicable in many cases.

Almost for all value of α, the proposed model (in both SA and SU forms) outperformed memory-based CF in terms of diversity and coverage. Memory-based CF, in average, gives about 15 and 20 percent lower diversity compared with SU and SA models, respectively. In terms of precision and novelty measures, for some values of α, memory-based CF outperformed the proposed model. However, there exist some values for α for which the proposed model outperforms CF in all evaluation measures. For example, in Netflix dataset, SA outperformed CF for the values of α in the range [0.75, 0.85] and SU outperformed CF for α > 0.98. According to the results, it can be said that CMR provides more accurate and diverse recommendations than memory-based models. However, SU compared with CMR has better performance in terms of the all metrics for $0.2 < α < 0.5$. Furthermore, for the range of $0.6 < α < 0.9$, SU outperforms CMR in terms of all evaluation metrics, while having almost the same novelty.

## 5. CONCLUSION

Recommendation systems are increasingly being used in various applications such as online stores. The aim of a recommender system is to use historical data about the users' behavior (e.g., their purchases as well as ratings on purchased items) and provide a list of items to each user such that they are likely to be purchased by the user in near future. Traditionally, performance of recommender systems has been evaluated by their precision. However, in order to find a practical application in real commercial systems, a recommender is needed to provide a recommendation list not only with high precision, but also with high levels of novelty and diversity. Often, the users would like to have novel and diverse recommendation list and such recommendations results in better user satisfaction.

In this manuscript, we introduced a probabilistic model, which has a full control on the level of precision, novelty and diversity. The model is a hybrid one as a linear integration of two state-automata based prediction models: specification maximizer and precision maximizer prediction models. The goal for specification maximizer model is to recommend items that are most specific to tastes of target users. The precision maximizer model aims at recommending items that have the highest probability to satisfy the target users. In order to construct the models, we took into account the sequence by which the items are purchased by the users. Our experiments showed that the proposed model could successfully deal with the dilemma of novelty-precision in recommendation systems and could result in better performance than classic recommenders such as user-based collaborative filtering and popularity-based algorithms.

## REFERENCES

[1]     G. Adomavicius and A. Tuzhilin, "Toward the next generation of recommender systems: a survey of the state-of-the-art and possible extensions," *IEEE Transs on Audio Electroacoustics Knowl and Data Eng,* vol. 17, pp. 734-749, 2005.

[2]     P. Resnick and H. R. Varian, "Recommender systems," *Communications of the ACM,* vol. 40, pp. 56-58, 1997.
[3]     X. Su and T. M. Khoshgoftaar, "A survey of collaborative filtering techniques," *Advances in artificial intelligence,* vol. 2009, p. 4, 2009.
[4]     M. Pazzani and D. Billsus, "Content-based recommendation systems," *The adaptive web,* pp. 325-341, 2007.
[5]     B. Sarwar, G. Karypis, J. Konstan, and J. Riedl, "Application of dimensionality reduction in recommender system-a case study," DTIC Document2000.
[6]     R. Burke, "Hybrid recommender systems: survey and experiments," *User Modeling and User-Adaptive Interaction,* vol. 12, pp. 331-370, 2002.
[7]     J. Konstan, B. Miller, D. Maltz, J. Herlocker, L. Gordon, and J. Riedl, "GroupLens: applying collaborative filtering in usenet news," *Communications of the ACM,* vol. 40, pp. 77-87, 1997.
[8]     B. Sarwar, G. Karypis, J. Konstan, and J. Riedl, "Item-based collaborative filtering recommendation algorithms," presented at the 10th international conference on World Wide Web, 2001.
[9]     A. Javari, J. Gharibshah, and M. Jalili, "Recommender Systems Based on Collaborative Filtering and Resource Allocation," *Social Network Analysis and Mining (to appear),* 2014.
[10]    S. Maslov and Y.-C. Zhang, "Extracting hidden information from knowledge networks," *Physical Review Letters,* vol. 87, p. 248701, 2001.
[11]    T. Hofmann, "Latent semantic models for collaborative filtering," *ACM Transactions on Information Systems (TOIS),* vol. 22, pp. 89-115, 2004.
[12]    J. S. Breese, D. Heckerman, and C. Kadie, "Empirical analysis of predictive algorithms for collaborative filtering," in *Proceedings of the Fourteenth conference on Uncertainty in artificial intelligence*, 1998, pp. 43-52.
[13]    G. Shani, R. I. Brafman, and D. Heckerman, "An MDP-based recommender system," in *Proceedings of the Eighteenth conference on Uncertainty in artificial intelligence*, 2002, pp. 453-460.
[14]    M. Deshpande and G. Karypis, "Item-based top-N recommendation algorithms," *ACM Transactions on Information Systems,* vol. 22, 2004.
[15]    D. F. J. Ben Schafer, Jon Herlocker and Shilad Sen, "Collaborative filtenng recommender systems," *LNCS: Lecture Notes In Computer Science,* vol. 4321, pp. 291-324, 2007.
[16]    G. Shani, D. Heckerman, and R. I. Brafman, "An MDP-based recommender system," *Journal of Machine Learning Research,* vol. 6, p. 1265, 2006.
[17]    B. Mobasher, H. Dai, T. Luo, and M. Nakagawa, "Using sequential and non-sequential patterns in predictive web usage mining tasks," in *Data Mining, 2002. ICDM 2003. Proceedings. 2002 IEEE International Conference on*, 2002, pp. 669-672.
[18]    A. Zimdars, D. M. Chickering, and C. Meek, "Using temporal data for making recommendations," in *Proceedings of the Seventeenth conference on Uncertainty in artificial intelligence*, 2001, pp. 580-588.
[19]    F. Garcin, C. Dimitrakakis, and B. Faltings, "Personalized News Recommendation with Context Trees," *arXiv preprint arXiv:1303.0665,* 2013.
[20]    J. L. Herlocker, J. A. Konstan, L. G. Terveen, and J. T. Riedl, "Evaluating collaborative filtering recommender systems," *ACM Transactions on Information Systems (TOIS),* vol. 22, pp. 5-53, 2004.
[21]    Ò. Celma and P. Herrera, "A new approach to evaluating novel recommendations," in *Recommendation Systems*, Lausanne, Switzerland, 2008, pp. 179-186.
[22]    S. M. McNee, J. Riedl, and J. A. Konstan, "Being accurate is not enough: how accuracy metrics have hurt recommender systems," in *CHI'06 extended abstracts on Human factors in computing systems*, 2006, pp. 1097-1101.
[23]    G. Adomavicius and Y. Kwon, "Improving aggregate recommendation diversity using ranking-based techniques," *IEEE Transactions on Knowledge and Data Engineering,* vol. 24, pp. 896-911, 2012.
[24]    T. Zhou, Z. Kuscsik, J.-G. Liu, M. Medo, J. R. Wakeling, and Y.-C. Zhang, "Solving the apparent diversity-accuracy dilemma of recommender systems," *Proceedings of the National Academy of Science of the United States of America,* vol. 107, pp. 4511-4515, 2010.
[25]    M. Deshpande and G. Karypis, "Selective Markov models for predicting Web page accesses," *ACM Transactions on Internet Technology (TOIT),* vol. 4, pp. 163-184, 2004.
[26]    R. R. Sarukkai, "Link prediction and path analysis using Markov chains," *Computer Networks,* vol. 33, pp. 377-386, 2000.
[27]    B. Mobasher, "Data mining for web personalization," in *The Adaptive Web*, ed: Springer, 2007, pp. 90-135.
[28]    A. L. Montgomery, S. Li, K. Srinivasan, and J. C. Liechty, "Modeling online browsing and path analysis using clickstream data," *Marketing Science,* vol. 23, pp. 579-595, 2004.

[29] Q. He, D. Jiang, Z. Liao, S. Hoi, K. Chang, E.-P. Lim, and H. Li, "Web query recommendation via sequential query prediction," in *Data Engineering, 2009. ICDE'09. IEEE 25th International Conference on*, 2009, pp. 1443-1454.

[30] H. Yu and M. O. Riedl, "A sequential recommendation approach for interactive personalized story generation," in *Proceedings of the 11th International Conference on Autonomous Agents and Multiagent Systems-Volume 1*, 2012, pp. 71-78.

[31] V. Nikulin, "OpenStudy: Recommendations of the Following Ten Lectures After Viewing a Set of Three Given Lectures," in *Proc. of ECML-PKDD 2011 Discovery Challenge Workshop*, 2011, pp. 59-69.

[32] R. Agrawal, S. Gollapudi, A. Halverson, and S. Ieong, "Diversifying Search Results," in *Web Search and Data Mining*, Barcelona, Spain, 2009, pp. 5-14.

[33] S. Vargas and P. Castells, "Rank and relevance in novelty and diversity metrics for recommender systems," in *Recomemnation Systems*, Chicago, Illinois, USA, 2011, pp. 109-116.

[34] C. Yu, L. Lakshmanan, and S. Amer-Yahia, "It takes variety to make a world: diversification in recommender systems," in *Proceedings of the 12th International Conference on Extending Database Technology: Advances in Database Technology*, 2009, pp. 368-378.

[35] C. Yu, L. V. S. Lakshmanan, and S. Amer-Yahia, "Recommendation diversification using explanations," presented at the IEEE International Conference on Data Engineering, 2009.

[36] S. Vargas and P. Castells, "Rank and relevance in novelty and diversity metrics for recommender systems," in *Proceedings of the fifth ACM conference on Recommender systems*, 2011, pp. 109-116.

[37] G. Shani and A. Gunawardana, "Evaluating recommendation systems," in *Recommender systems handbook*, ed: Springer, 2011, pp. 257-297.

[38] C.-N. Ziegle, S. M. McNee, J. A. Konstan, and G. Lausen, "Improving recommendation lists through topic diversification," in *World Wide Web*, Chiba, Japan, 2005, pp. 22-32.

[39] Z.-K. Zhang, T. Zhou, and Y.-C. Zhang, "Personalized recommendation via integrated diffusion on user–item–tag tripartite graphs," *Physica A: Statistical Mechanics and its Applications,* vol. 389, pp. 179-186, 2010.

[40] S. Rendle, C. Freudenthaler, and L. Schmidt-Thieme, "Factorizing personalized Markov chains for next-basket recommendation," in *Proceedings of the 19th international conference on World wide web*, 2010, pp. 811-820.

[41] A. Javari and M. Jalili, "Accurate and novel recommendations: An algorithm based on popularity forecasting," *ACM Transactions on Intelligent Systems and Technology (to appear),* 2014.